\documentclass[journal]{journal}

\usepackage{algorithmic}
\usepackage{graphicx}
\usepackage[caption=false]{subfig}

\usepackage{textcomp}
\usepackage{xcolor}
\usepackage{float}
\usepackage{gensymb}
\def\BibTeX{{\rm B\kern-.05em{\sc i\kern-.025em b}\kern-.08em
    T\kern-.1667em\lower.7ex\hbox{E}\kern-.125emX}}
    
\begin{document}
%
\title{``Vironment'':
\\
An Art of Wearable Social Distancing}
%
%
%

\author{Steve~Mann,
Cayden~Pierce,
Christopher~Tong,
Christina~Mann
\thanks{We wish to thank AMD, SwimOP, and Swim Drink Fish}
}

\maketitle
\thispagestyle{empty}

\begin{abstract}
``Vironment'' is a series of art pieces, social commentary, technology, etc.,
based on wearable health technologies of social-distancing,
culminating in a social-distancing device that takes the familiar world of
security and surveillance technologies that surround us
and re-situates it on the body of the wearer (technologies that become part of us).
This piece also introduces a conceptual framework for (1) the sensing of the self together with (2) sensing of others and (3) sensing of the environment around us.
\end{abstract}


%
\IEEEpeerreviewmaketitle

%
%
%
%

\begin{figure}
\includegraphics[width=\columnwidth]{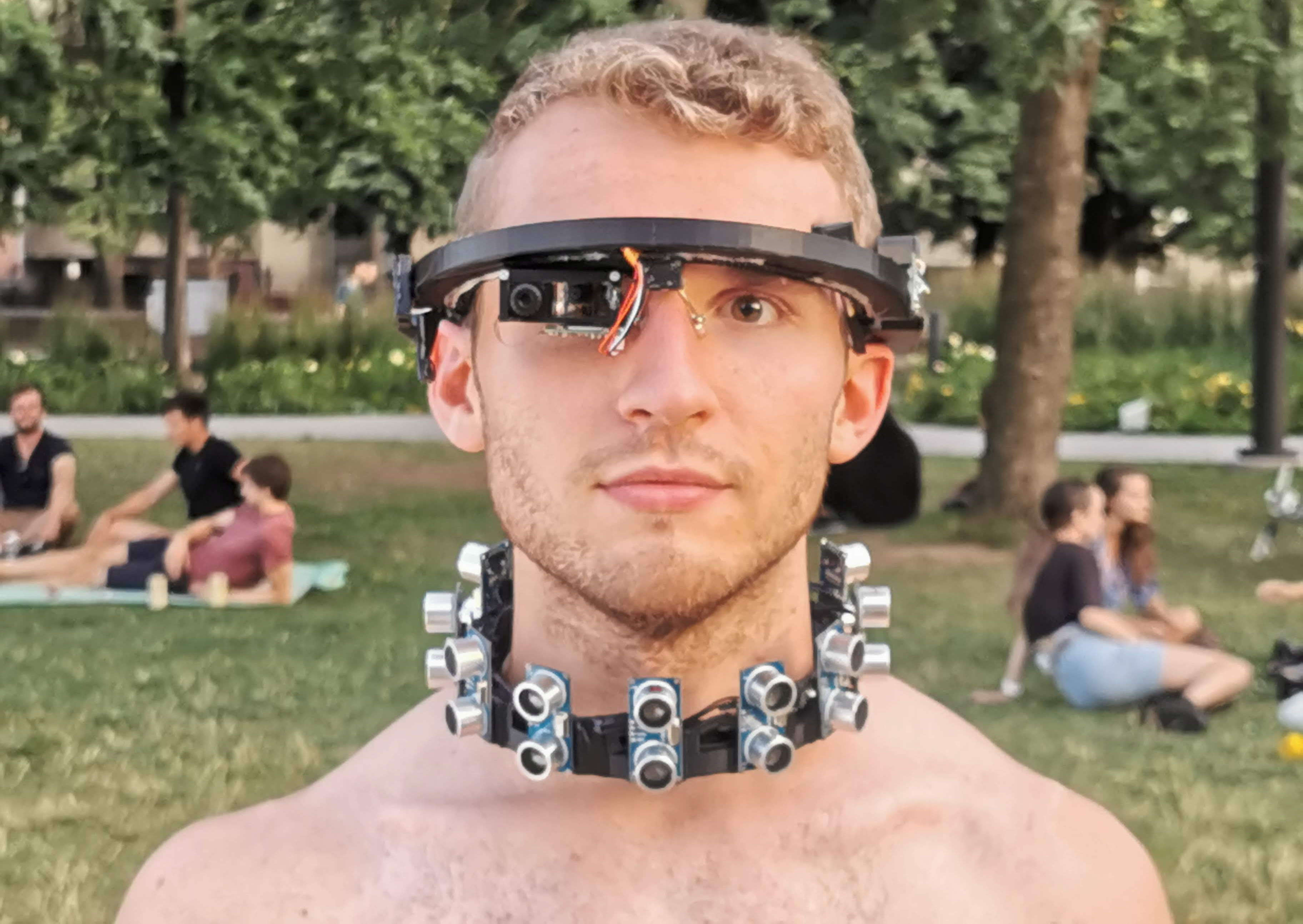}
\caption{3D printed version of Vironment 3.0 with rigidly aligned sensors and OpenEyeTap computer vision and display system.
}
\label{fig:vironmen3}
\end{figure}

\section{Background and Introduction}
\IEEEPARstart{I}{nvironment}
is a new concept defined in contrast to the environment.  Whereas the environment is our surroundings, the invironment is us, ourselves, and in particular, includes that part of us that we consider to be ourselves, e.g. our shoes, clothing, eyeglasses, and the like, along with a certain space, or {\em social-distance} around us, often defined in the context of wearable computing\cite{wearableai, mannieeecomputer}.  The concept of invironment is of particular relevance in the era of health pandemics and pandemic awareness as global society asks questions about health, safety, wellness, and the individual's right to utilize technologies to assist in these areas\cite{nayak2020changing, amft2020wearables, nestor2021dear, linares2021physician}. Technologies for health and wellness are rapidly becoming viewed as extensions of one's body, which raises questions about human rights to technology, the human condition, and the extension of ``the self~\cite{fitzgerald1963self}'' into the space around the body. Specifically, with the introduction of social-distancing, questions arise regarding the area in which ``the self'' exists and where ``the self'' ends.

In further inquiry of these concepts, a series of interactive art pieces are constructed to highlight the good and bad of social-sistancing, i.e. some of its benefits, downsides, and absurdities of social distancing, based on the idea of a social-distancing necklace, inspired by the commonly worn spiked necklace shown in Fig.~\ref{fig:normal_spike_necklace}. We develop wearable computing systems which promote and consider the health practice of social distancing from a health and wellness perspective, a realistic sociological perspective, and a technological perspective.

Surveillance is well-known in the areas of
smart buildings, smart streets, smart cities, etc. from an ``Internet of Things'' (IoT) perspective.  This is the traditional situation in which sensors
exist in the environment around us. More recently, ``wearables'' (wearable technology) has emerged as a new discipline in which sensors are affixed to people rather than things~\cite{barfieldbook2, starner97, mannieeecomputer}, giving rise to ``WearableAI\cite{wearableai}'' (Wearable Artificial Intelligence'').
What is most important about wearable technology is not so much the proximity to the body, but, more importantly, the ability for this technology to function as an agent of the mind and body's freewill and self-determination over own one's own destiny, i.e. 
``sousveillance''\cite{mann2002sousveillance, mann2003sousveillance, fletcher2011, measuringVeillanceVixels, michael2012sousveillance, bakir2010sousveillance, Bradshaw2013Police, Freshwater2013Revisiting} (inverse surveillance, sensors-on-people) and self-sensing e.g. Quantified Self-Sensing (QSS)\cite{intelligentimageprocessing, kelly2007quantified}, as outlined in Fig 1 of the ITTI paper\cite{itti}.

Here, we present a series of art pieces, social commentary, technology, inventions, etc. which raise questions deeply connected to the paradigm shifts of social-distancing and wearable computing and how they interact.

\section{Self and Technology, Society, and Environment}
In the past cities were more important than countries in terms of boundaries, e.g. walled cities thousands of years ago.  Next countries emerged as important.  Finally in the era of global pandemics, we're seeing world governance, and the reduced autonomy of countries.  What matters now is clothes, i.e. our individual selves become in some sense the boundary of greatest importance.  So in regards to ``crossing borders'', we've witnessed the evolution from cities, to countries, to clothes.

Vironment is an exploration of the following three elements:
\begin{itemize}
    \item Self and technology (e.g. the combination of human and machines, ``cyborg'', ``humachine'', augmented human, etc.);
    \item Self and society (interaction between humans, augmented humans, etc.);
    \item Self and the environment (interaction between the augmented human and the natural or built environment, e.g. cyborg-city interaction, etc.)\cite{mann2020sensing}.
\end{itemize}

\section{Vironment 1.0/Social-Bubble}
Vironment is a series of art installations, design interventions, and inventions aimed at understanding and reconstructing social distance as the sometimes soft and fuzzy boundary between the invironment and the environment. Through these creations arises a deeper inquiry into the relationship between the body and the spacial sphere around the body, which together form much of the concept of ``self''. The design pieces presented here explore the extension of the self into the environment, an area of space which has become of utmost social significance in the era of health pandemics as individuals individuate and define themselves by the ``social bubble'' that exists around them.

\begin{figure}
    \centering
    \includegraphics[height=4in]{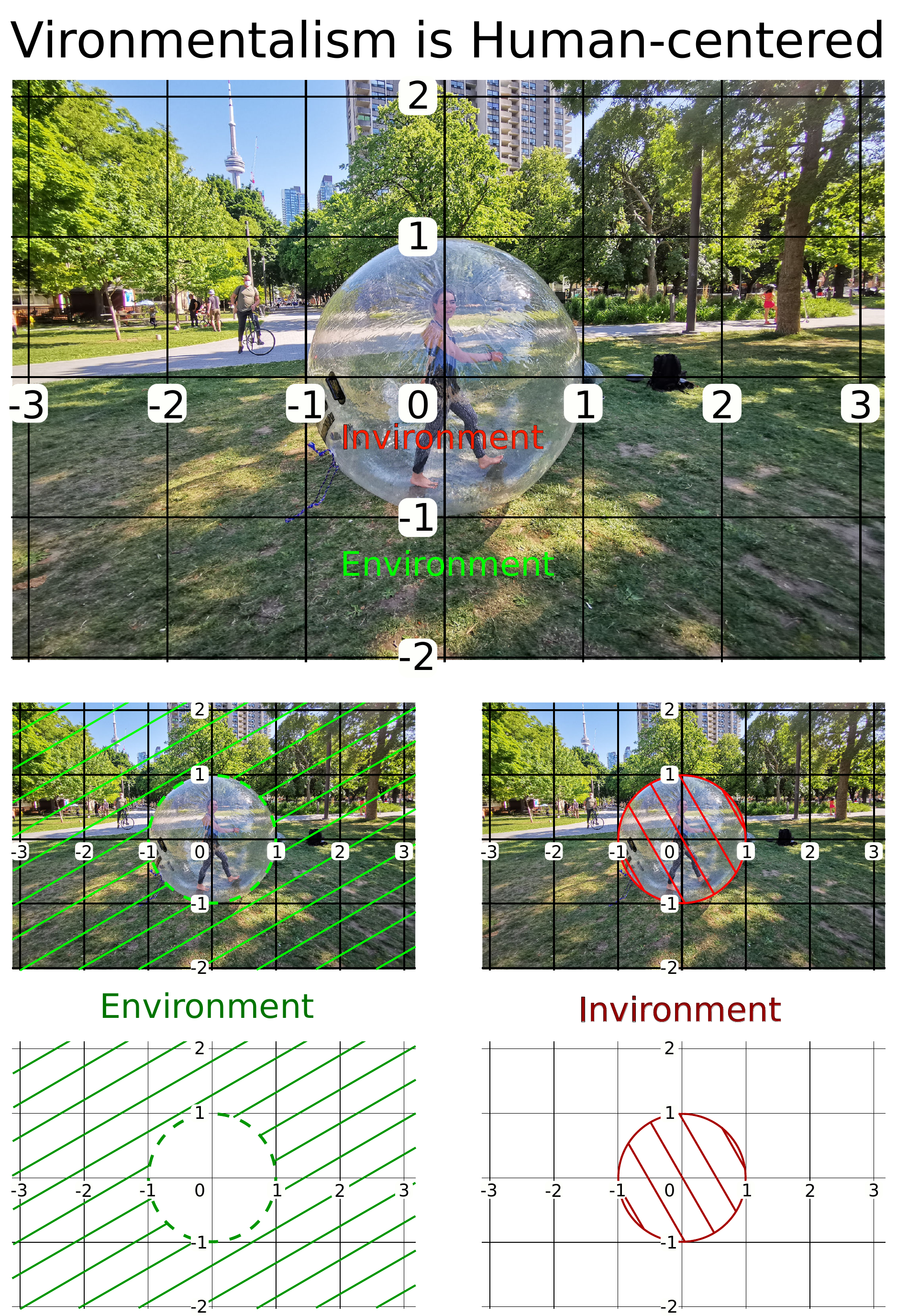}
    \caption{``Vironment/Social-Bubble'', or simply ``Vironment 1.0'' is a social commentary on social-distancing to help us rethink the natural and social environment and our natural and social ecosystems in times of pandemics.}
    \label{fig:vironment}
\end{figure}

\begin{figure}
    \includegraphics[width=\columnwidth]{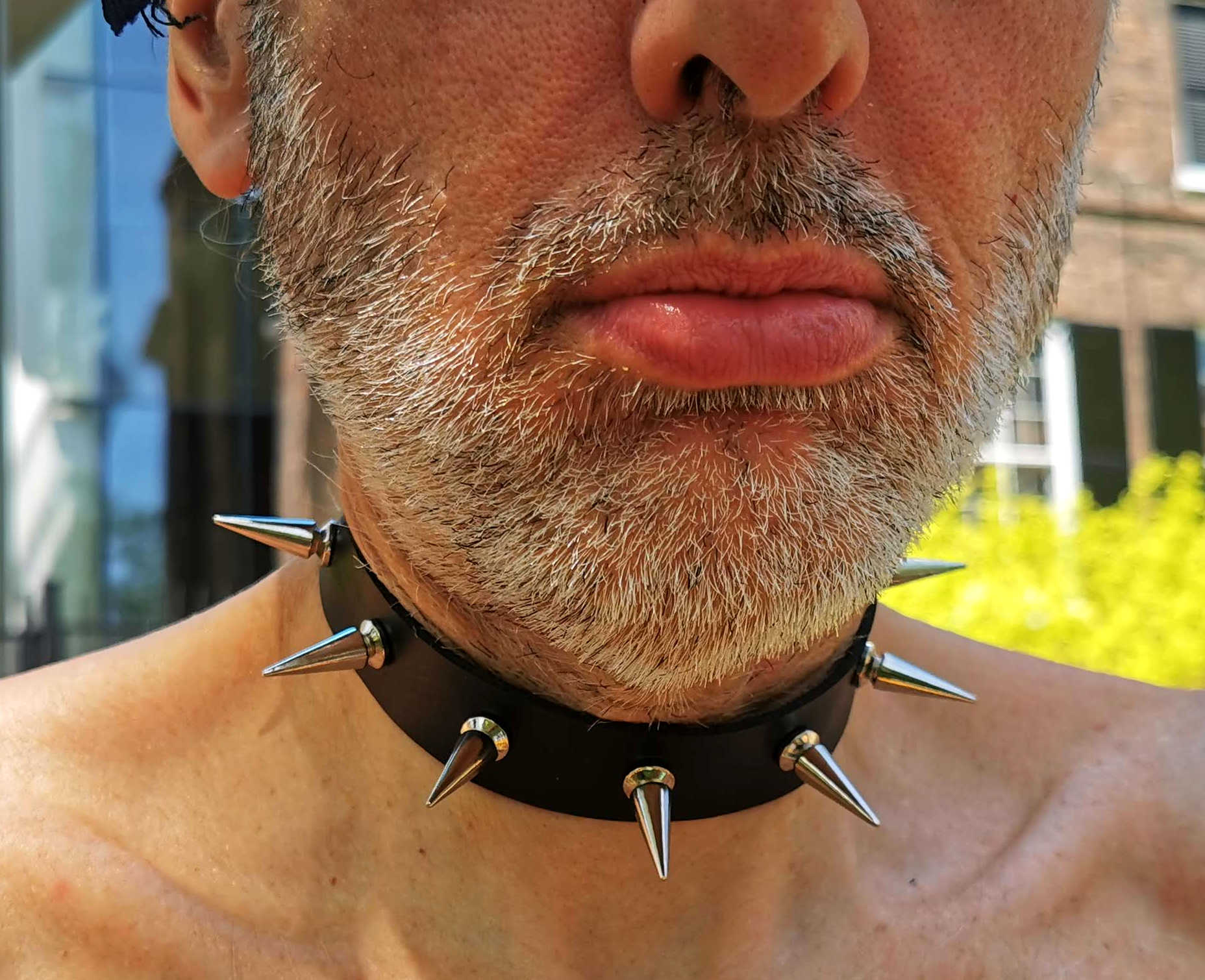}
    \caption{Inspiration for Vironment 2.0.}
    \label{fig:normal_spike_necklace}
\end{figure}

One piece we name ``Vironment/Social-Bubble", or, simply, ``Vironment 1.0", is shown in Fig~\ref{fig:vironment}.  
It is based on the use of a ball commonly used as a water-based amusement ride, called the ``waterball ride'' or ``water ball ride''.
In the context of a vessel surrounding the rider, when used on water, it represents an example of the interaction between humans, water, and technology.  (The field of ``WaterHCI = Water-Human-Computer Interaction'' originated in Ontario in the 1960s and 1970s, and relates to the concept of ``Fluidic User Interface''\cite{mann2005fl}).

Here we explore the use of such a ball anywhere, such as on land, and not just on water only.

Vironment 1.0 takes the idea of social-distancing
quite literally.  Many of the parks in our neighbourhood post signs warning
us to keep our social distance of two metres or six feet.  Vironment 1.0 is a two metre diameter plastic bubble in which a participant is placed, and the bubble is
filled through a HEPA (High Efficiency Particulate Air) filter connected to
a blower supplied by a solar-powered battery system.
The filter and blower assembly was constructed from an automobile heater blower and runs on a small 12 volt battery which is charged by solar power.

Two individuals crossing paths will be forced to maintain a proper 2m social distance. This demonstrates an ideal situation for an individual wishing to avoid contact with pandemic diseases by providing the wearer with clean, filtered air and a visible ``social bubble" barrier. Also demonstrated is the inherent impossibility of social-distancing in some situations, as the individual in the bubble cannot fit through doorways,
or even walk down a narrow sidewalk whilst encapsulated in an impenetrable social bubble.

This piece (Vironment 1.0) raises questions about what space ``belongs" to an individual. Does an individual encompass only their brain? Their whole body? The brain, body, and the clothing on the body, and maybe also some of the space that surrounds the body and its clothes?  This is a vital question to ask in times when the health and well-being of individuals and society is predicated on everyone existing in and maintaining their own invironmental space.  Vironment 1.0 serves as a visceral example and visualization of the social space that has become normal (and often mandatory) in the era of health pandemics and pandemic awareness.

\section{Vironment 2.0}
Vironment 2.0 borrows from a theatre of the absurd and the d\'{e}tournement (Situationist)
visual art tradition, as something so hideous as to hopefully make itself unnecessary once it has raised the awareness needed to make it unnecessary.  It is important to note that spiked chokers are not very socially acceptable to begin with.  Here the location of this wearable compliments the off-putting aesthetic\cite{zeagler2017wear}.

\begin{figure}
    \includegraphics[width=\columnwidth]{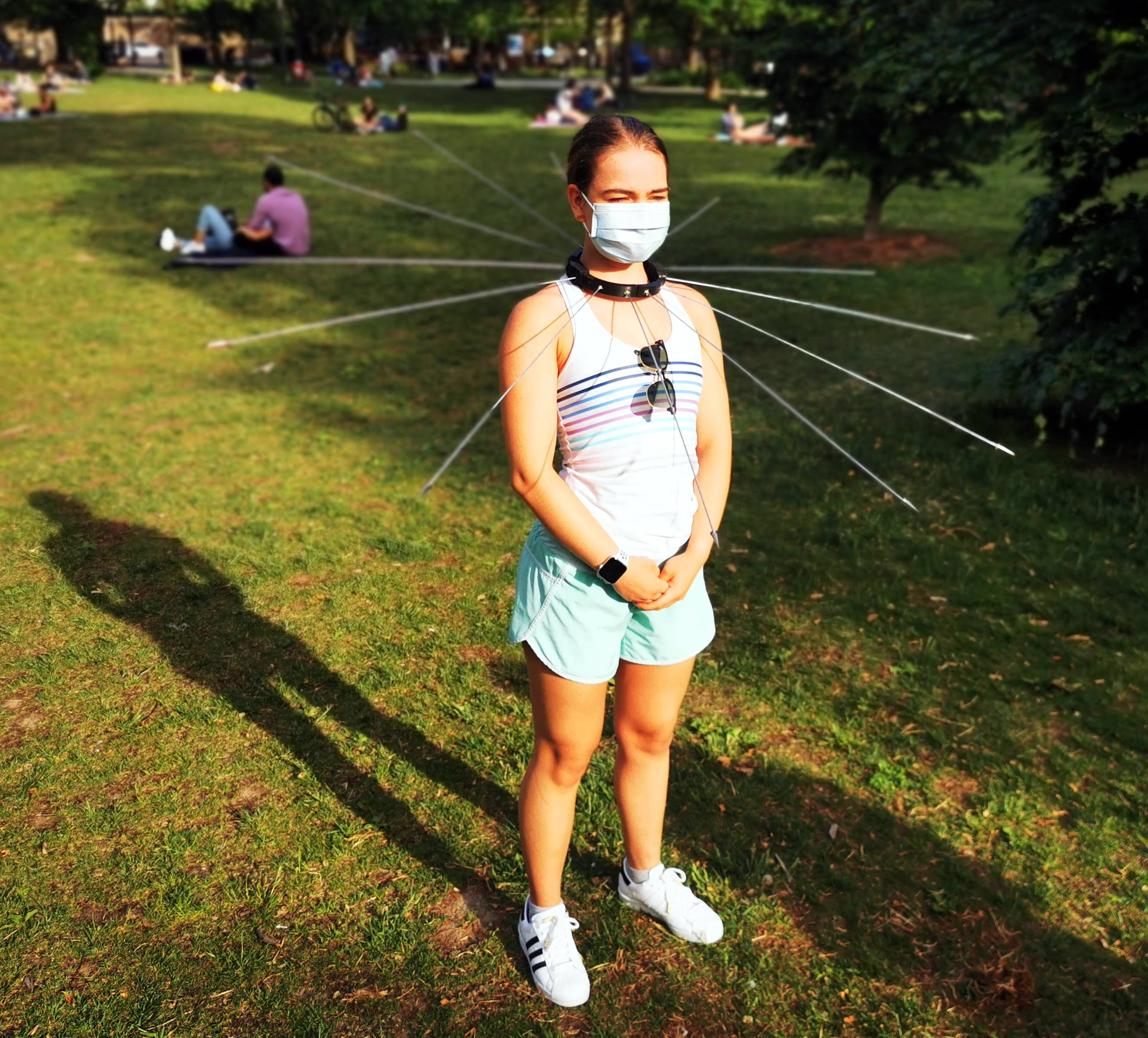}
    \caption{Vironment 2.0: Social distancing spikes.}
    \label{fig:longspikes}
\end{figure}

Vironment 2.0 is an extension of the common spiked necklace, in which the spikes are simply made longer, as shown in Fig.~\ref{fig:longspikes}. The necklace was 3D printed and the spike extensions were cut from steel wire, tipped with the same spikes as in a common spiked necklace.

This piece more assertively defines the space around oneself. Instead of simply blocking out anyone from entering into one's social bubble, Vironment 2.0 threatens consequences for anyone entering into one's social bubble. This touches on the real-world health risks associated with entering into others' social bubbles, and the physically detrimental ramifications that can be dealt to those who disrespect maintenance of the social bubble.

\section{Vironment 3.0}
\begin{figure}
\includegraphics[width=\columnwidth]{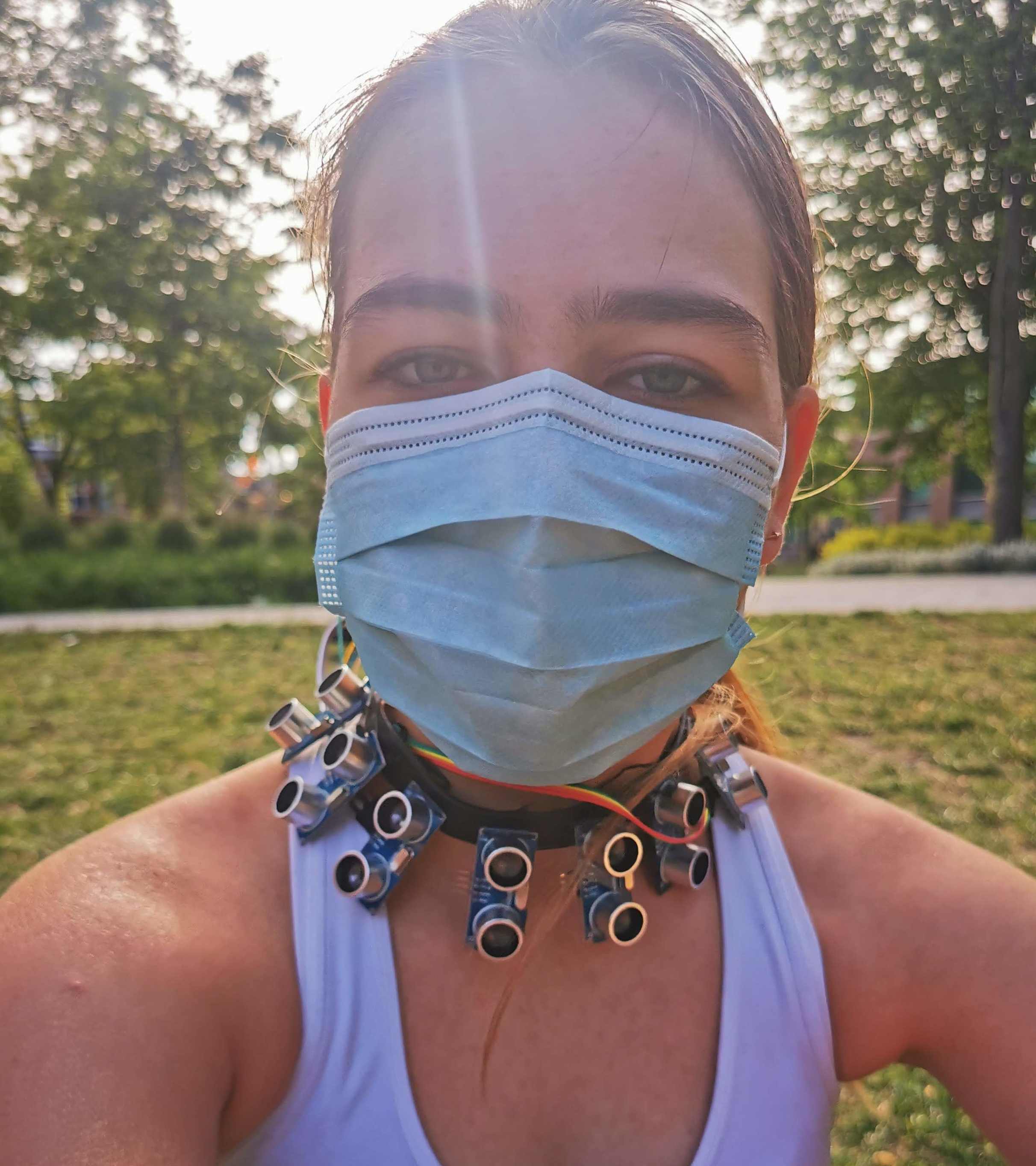}
    \caption{Vironment 3.0 with softband.}
\label{fig:softband}
\end{figure}

Further, we present a virtual version of Vironment/Invironment, see Fig~\ref{fig:vironmen3}, Fig~\ref{fig:softband}, and Fig~\ref{fig:necklace}.
The virtual Vironment uses sonar to alert the wearer in regards to violation of
social-distancing.  In particular, we use 12 HC-SR04 Ultrasonic sensors which each have a detectable range from 2cm to 4 metres (approx. 13.1 feet).  These 12 sensors are arranged around a circle, at 30 degree intervals, analogous to a clock face, i.e. at the cardinal 12-hour clock face directions and worn around the neck of the individual. Vironment 3.0 is a distance sensor with a 360\degree\ veillance field surrounding the wearer. A blinding bright LED will light up and a car horn (worn in a backpack) will sound whenever this apparatus detects an individual entering into the 2m social radius existing around the wearer. This serves as a barrier-less and non-violent approach to social-distancing that maintains the importance of the social bubble, while maintaining a notification for the wearer and the individual imposing upon the wearer's personal space.


\begin{figure}
\includegraphics[width=\columnwidth]{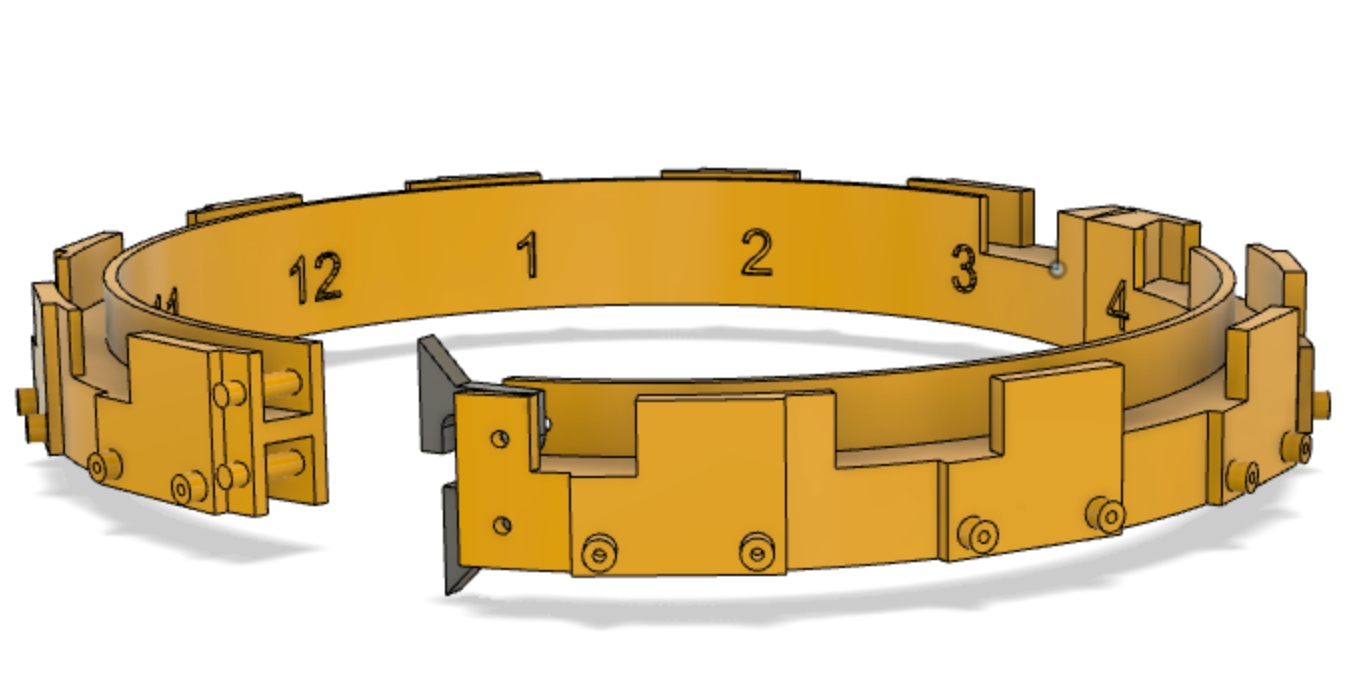}
    \caption{CAD of Vironment 3.0 made in Fusion 360 for 3D printing}
\label{fig:sonar_social}
\end{figure}

\begin{figure*} 
\centering
\subfloat{\includegraphics[height=0.6\columnwidth]{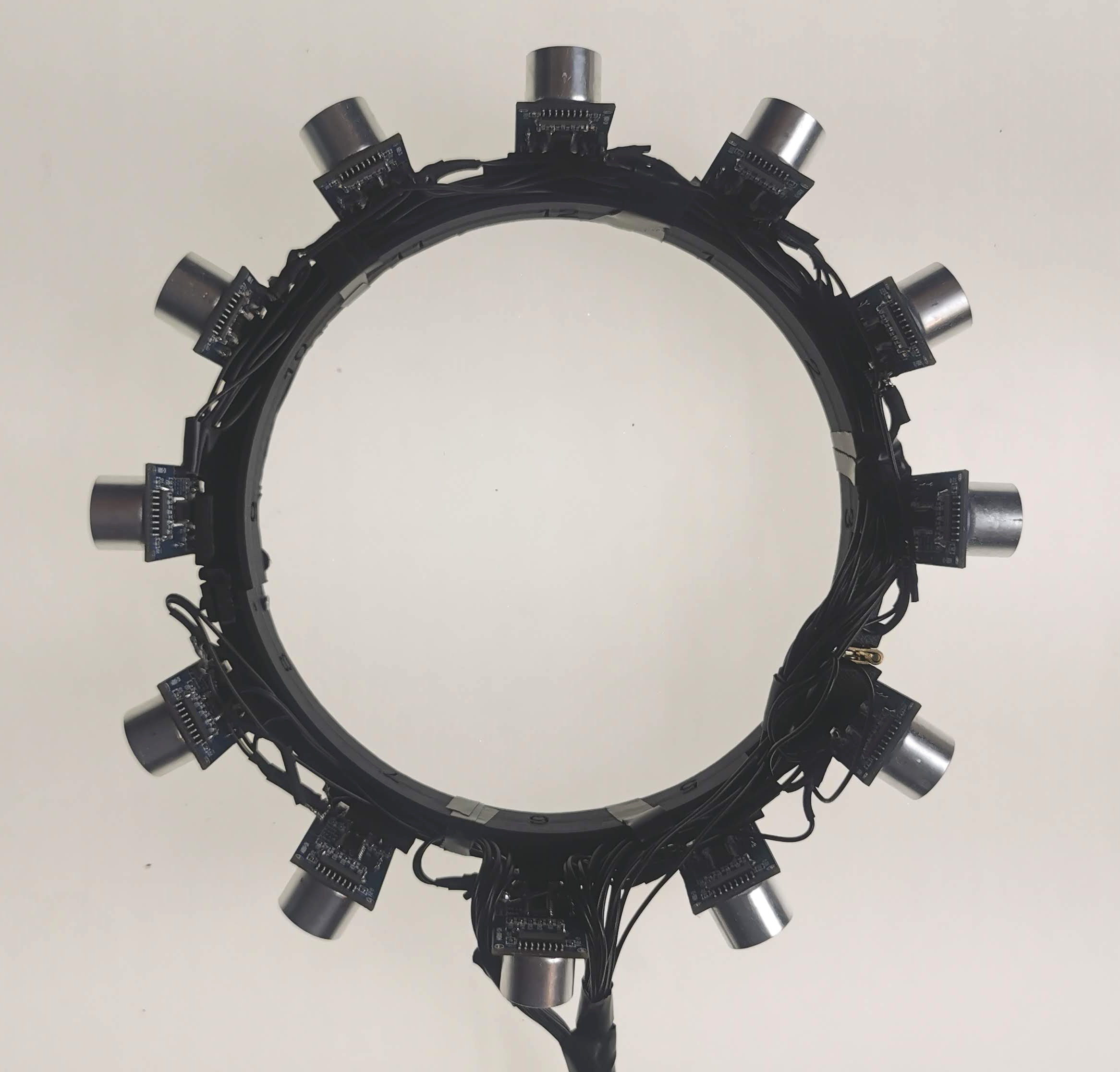}}
\hfil
\subfloat{\includegraphics[height=0.6\columnwidth]{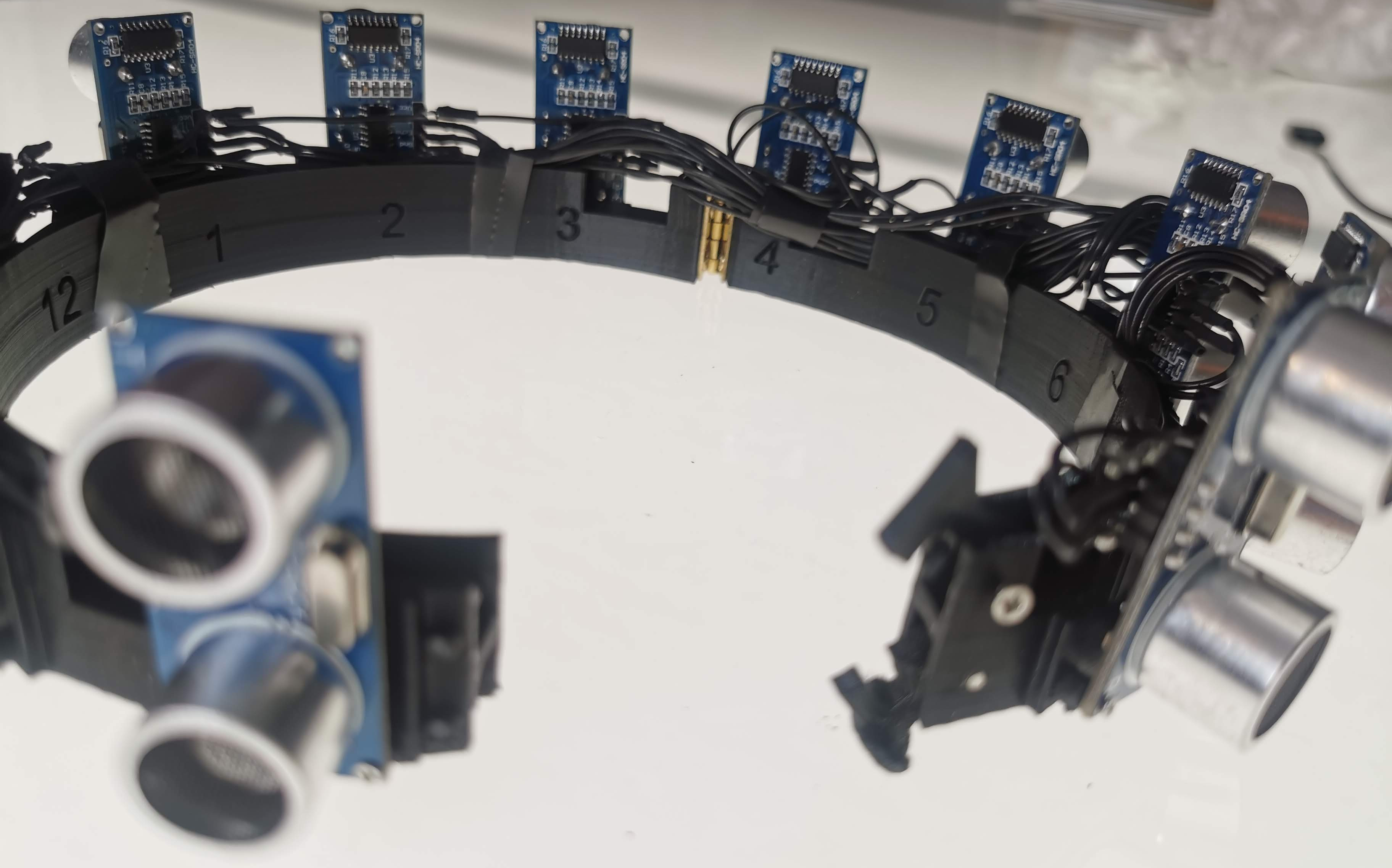}}
\caption{The Vironment 3.0 necklace with hardband.}
\label{fig:necklace}
\end{figure*}



A 3-D printed version of Vironment 3.0 was made in which the necklace is hollowed out in order to route wires for the sonars more discretely. The circuit board was also moved to the back of the necklace to act as a counterweight. In previous designs the necklace was not level, with the forward facing sonars pitched down.  Adding a counterweight allows the necklace to align in parallel with the ground, ensuring the sensors can sense the presence of any imposing individuals within a 360\degree\ field around the wearer. 

In order to increase the ease and speed at which the apparatus is mounted on the neck, a hinge was placed to connect the two 3D printed parts of the necklace, allowing it to open and close for easy wearability. A double latch system was designed and included as a lock to keep the necklace shut around the user's neck. See Fig~\ref{fig:sonar_social} for a CAD model of the Vironment 3.0 piece.

The electrical system is composed of 12 HC-SR04 Ultrasonic sensors, with all Vcc pins on a 5 Volt bus, and all Gnd pins on a Ground bus. Each sensor has two data wires, one ``Trigger", which triggers the ultrasonic wave, and one `Echo", which outputs information regarding the time delay (and thus, distance measured). Because there are 12 sonars each with two data lines (24 lines), four MC14051B analog multiplexers are used to switch between the sonars, with 2 multiplexers chained for ``Trigger"s
from the sensors, and 2 multiplexers chained for ``Echo"s. The system is controlled by an Espressif ESP-32 DevKit C microcontroller, which controls the multiplexers, provides power to the sensors, and processes the data. These few components are mounted and soldered to a piece of perfboard. The necklace interfaces with the display system (the EyeTap) using a direct USB connection.

In the next, future iteration of Vironment, Vironment 4.0, instead of 12 pairs of ultrasound transducers, there will be 12 individual transducers in which each is both a transmitter and a receiver. This will decrease the weight of the wearable as well as its height (the double stacked sonar hits the user's chin), from about 48mm to 16mm, i.e. about one third the height of Vironment 3.0.


\subsection{PPI (Plan Position Indicator)}
\begin{figure}
    \centering
    \includegraphics[height=1.85in]{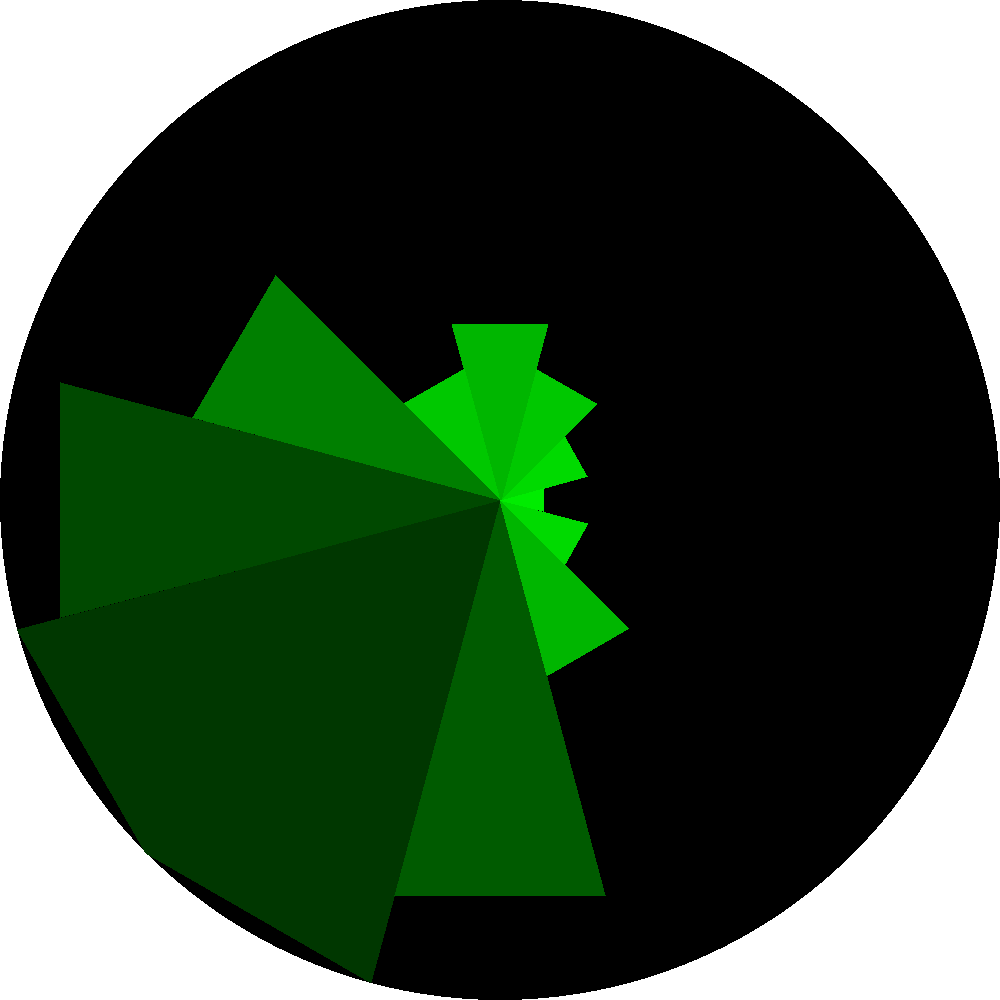}
    \caption{The P.P.I. live user interface representing the social bubble around the user. The distance of the dodecant boundary to the center represents the distance of another individual to the wearer, at that specific location. The color is also modulated from black (infinite) to bright green (2cm) to represent the distance.}
    \label{fig:ppi}
\end{figure}

Finally, we combine two concepts:
(1) the concept of the PPI (Plan Position Indicator) from
radar~\cite{harold1946plan} (see Fig~\ref{fig:ppi}),
and (2) the use of a clockface, especially in shooter-based reconnaissance
where a shooter is instructed as to the location of enemy targets in reference to a clock face. The necklace was updated to reflect the layout of a clock, with a sonar placed at every hour position. In egocentric coordinates, the sonar at 12 o'clock points forward and the sonar at 6 o'clock points backwards. The numbers of the clock are also engraved on the inside of the collar right behind their respective sonars, to aid the user in putting the necklace on properly. 

As such, rather than quadrants or octants, we use dodecants, dividing the space into 12 equally sized sectors.

This layout of PPI and clockface is displayed live on AR glasses worn by the user. The AR view is achieved by a microdisplay UI running on an EyeTap \cite{mann2002eyetap} smart glasses display, as shown in Fig~\ref{fig:vironmen3}.  The processor on the necklace (an ESP-32) communicates sonar information directly to the EyeTap's on-board Raspberry Pi Zero W Single Board Computer (SBC). 

Thus, the wearer of Vironment 3.0 can visualize (in AR) social distancing information using a Natural User Interface (NUI).

\begin{figure*} 
\centering
\subfloat{\includegraphics[height=0.72\columnwidth]{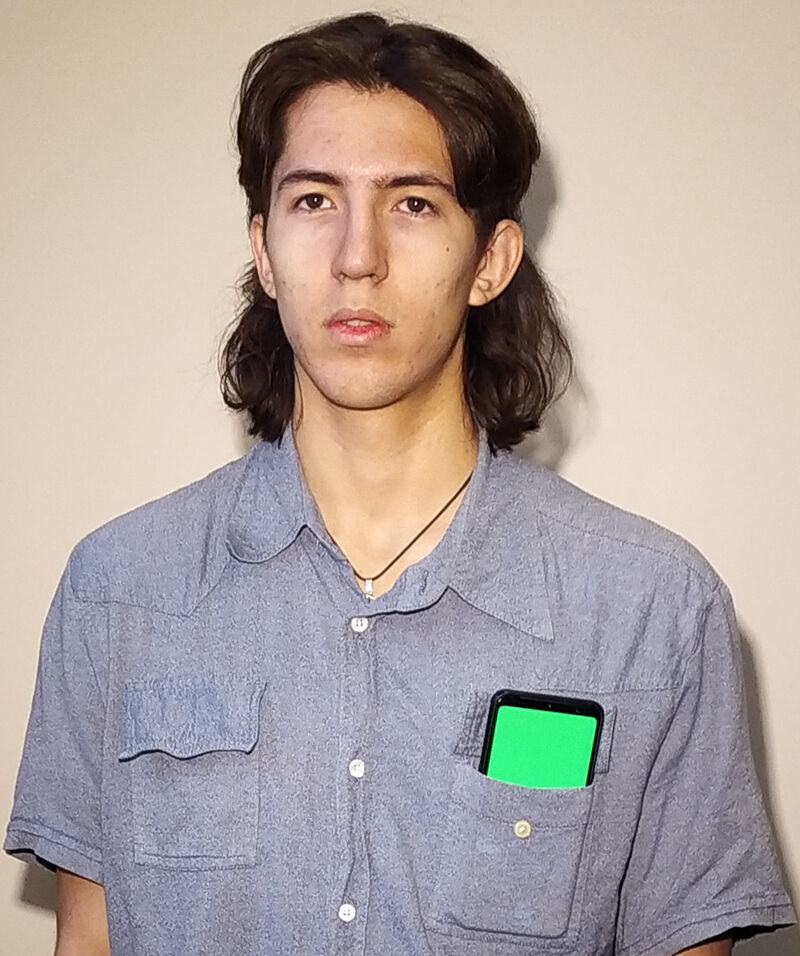}}
\hfil
\subfloat{\includegraphics[height=0.72\columnwidth]{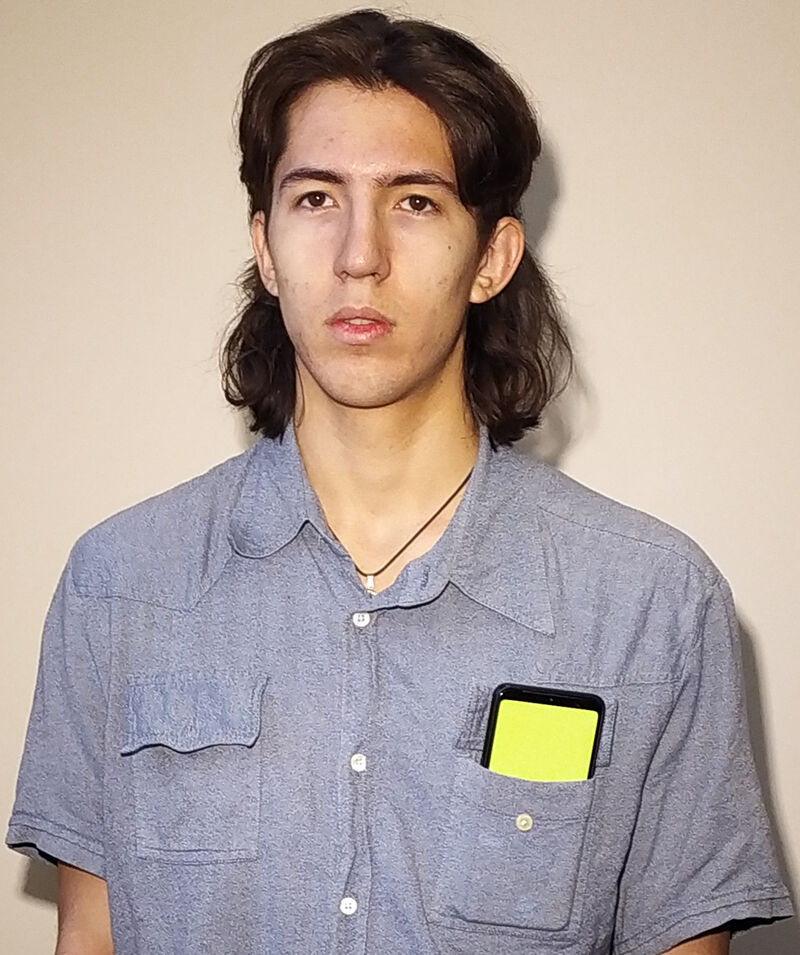}}
\hfil
\subfloat{\includegraphics[height=0.72\columnwidth]{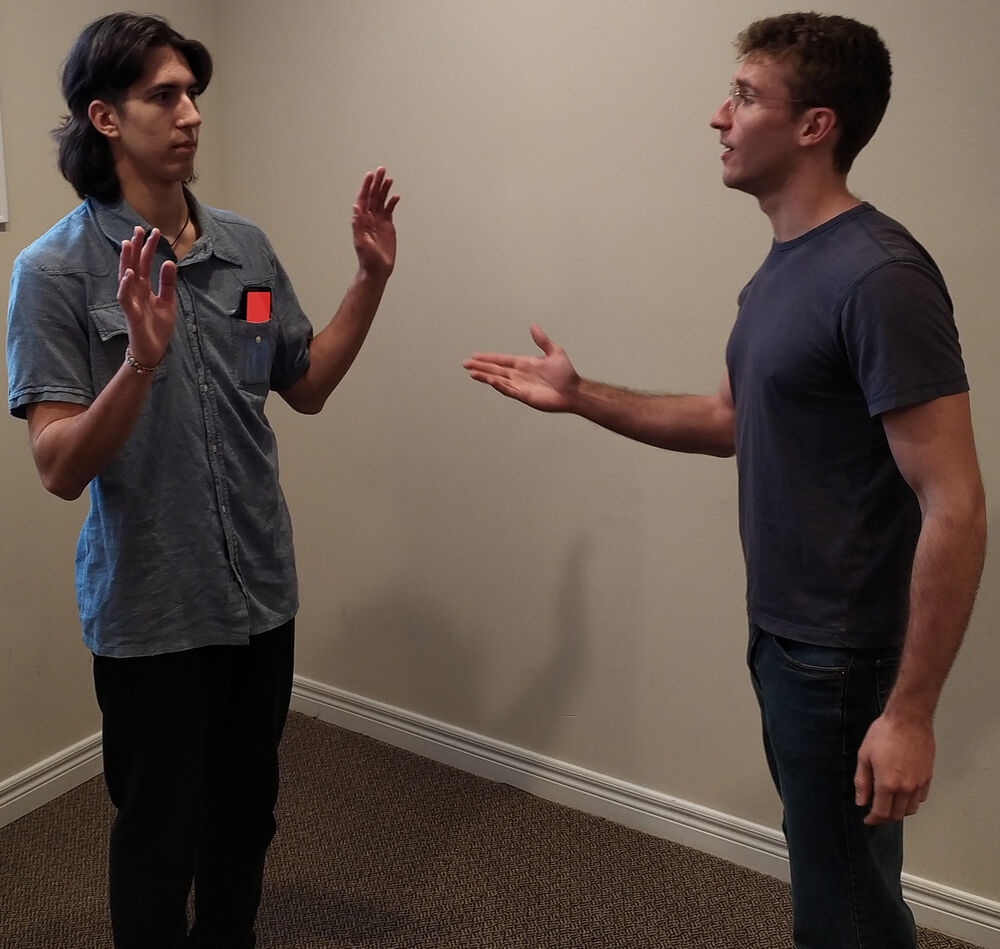}}
\caption{Social Distancer app: A wearable indicator for social distance. The application is run on a smartphone in the shirt pocket (with the screen facing out). The screen is green if proper social distance is maintained, yellow if on the verge of danger, and red if social distance is compromised.}
\label{fig:mobilesocialdistancer}
\end{figure*}

\section{Social Distancer app}
Many of the widely used and discussed social distancing technologies are surveillant - they rely on a centralized authority collecting data on citizens in order to maintain a social distancing and contact-tracing database. ``Vironment'' is different in its ability to encourage, or enforce, social distancing without the requirement of surveillant systems.

To continue exploration in this direction, we developed a social distancing smartphone app with the ability to be immediately deployed to billions of people. It runs on consumer hardware and presents a means of social distancing that is sousveillant, open, and available to everyone to use without extraneous hardware and without reliance on any central authority. The Social Distancer app can be seen in Fig~\ref{fig:mobilesocialdistancer}.


The Social Distancer app runs on a smart phone equipped with a forward facing camera. The application collects a live stream of video data from the front facing camera which it runs through an object detection neural network (MobileNet pretrained on Coco dataset \cite{DBLP:journals/corr/abs-1801-04381}). A filter is placed on the output of the network to only accept predictions of people in the image frame (ignoring recognition of other objects). The bounding box around the person is combined with the average height of a person ($\sim$1.65 meters \cite{ncd2016century}) to estimate the distance between the Social Distancer app and the individual being detected by the camera. The screen of the Social Distancer app is green when social distancing is maintained, yellow when individuals are 6 to 7 feet away (almost breaking social distancing), and red when social distancing has been compromised (i.e. another individual has intruded into the wearer's inviroment).

The user of the system places the Social Distancer app in the breast pocket of their clothing with the screen and front camera facing outwards. In this configuration, the Social Distancer app serves as a way to remind others to maintain a healthy and safe distance.








\section{Conclusion}
We have introduced a series of wearable health systems collectively named ``Invironment''.
Vironment 1.0, and 2.0 were forms of social commentary highlighting design
questions which were answered through a social-distancing sonar system,
Vironment 3.0.  Vironment 3.0 consists of an array of sonar sensors arranged in a circle at 30 degree intervals, i.e. as the 12 cardinal directions of the hours of a 12-hour clock face.  In this system we designed a PPI (Plan Position Indicator) on an EyeTap
display to show the wearer a top-down ``map'' of their risk space in regards to
violators of social distance. Finally, the Social Distancer app is a deployable, sousveillant system which promotes social distancing while avoiding the use of centralized surveillant data tracking.


\appendices
\section*{Acknowledgment}

The authors would like to thank Kyle Simmons for his feedback on the electrical design of the sonar social distancer ``Vironment 3.0" system.

\ifCLASSOPTIONcaptionsoff
  \newpage
\fi



%
\IEEEtriggeratref{21}
\bibliographystyle{IEEEtran}
\bibliography{chirplet}


%








\end{document}